\documentclass[conference]{./style/IEEEtran2/IEEEtran}
\usepackage{cite}
\ifCLASSINFOpdf
\else
\fi
\usepackage{amsmath}
\usepackage{listings}
\usepackage{adjustbox}
\usepackage{xspace} 
\usepackage{graphicx}
\usepackage{fancyvrb}
\usepackage{float}
\usepackage{pdfpages}
\usepackage[english]{babel}
\usepackage{multirow}
\usepackage{array}
\usepackage{color}
\usepackage{soul}
\usepackage{url}
\usepackage{hyperref}
\addto\extrasenglish{%
}
\usepackage{colortbl}
\usepackage[font=footnotesize]{caption}
\usepackage{subcaption}
\usepackage{setspace}
\usepackage{soul}
\usepackage{./style/slashbox/slashbox}
\usepackage{./style/bibspacing} 

\RequirePackage[normalem]{ulem}
\RequirePackage{color}\definecolor{RED}{rgb}{1,0,0}\definecolor{BLUE}{rgb}{0,0,1}

\begin{document}

\bstctlcite{IEEEexample:BSTcontrol}

\title{Anomaly Detection in High Performance Computers: A Vicinity Perspective}

\author{
\IEEEauthorblockN{Siavash Ghiasvand\IEEEauthorrefmark{2} and Florina M. Ciorba\IEEEauthorrefmark{1}}
\IEEEauthorblockA{
\IEEEauthorrefmark{2}Technische Universit\"at Dresden, Germany\\
\IEEEauthorrefmark{1}University of Basel, Switzerland
}
}

\makeatletter
\def\paragraph{\@startsection{paragraph}
				 	{5}
					{\z@}
	                                 {3.25ex \@plus1ex \@minus.2ex}
        		                        {\z@}
                  		       {\normalfont\normalsize\bfseries}
		       }
\makeatother
\maketitle
\begin{abstract}
In response to the demand for higher computational power, the number of computing nodes in high performance computers (HPC) increases rapidly.
Exascale HPC systems are expected to arrive by 2020.
With drastic increase in the number of HPC system components, it is expected to observe a sudden increase in the number of failures which, consequently, poses a threat to the continuous operation of the HPC systems.
Detecting failures as early as possible and, ideally, predicting them, is a necessary step to avoid interruptions in HPC systems operation.
Anomaly detection is a well-known general purpose approach for failure detection, in computing systems.
The majority of existing methods are designed for specific architectures, require adjustments on the computing systems hardware and software, need excessive information, or pose a threat to users' and systems' privacy.
This work proposes a node failure detection mechanism based on a vicinity-based statistical anomaly detection approach using passively collected and anonymized system log entries.
Application of the proposed approach on system logs collected over 8 months indicates an anomaly detection precision between 62\% to 81\%.
\end{abstract}

\IEEEpeerreviewmaketitle

\section{Introduction}
\label{sec:introduction}

In response to the demand for higher computational power, the number of components in high performance computers (HPC) rapidly increases~\cite{TOP500Cores}.
It is expected that Exascale HPC systems become available by 2020~\cite{Service618}.
Besides increasing the quantity of computing resources, achieving high performance is also dependent on the optimized utilization of available resources, as well as on the continuous operation of the HPC system as a whole.
Over the past decades, scientists proposed various methods and algorithms to adjust the workload on HPC systems to achieve the highest possible performance.
With the drastic increase in the number of HPC system components, it is expected to observe a sudden increase in the number of failures which consequently poses a threat to the continuous operation of the HPC systems~\cite{cappelloExascaleResilience20142014}.
Therefore, detecting failures as early as possible and, ideally, predicting them, is a necessary step to avoid interruptions in continuous HPC systems operation.

A failure in general is an (observed) incorrect behavior with respect to the expected behavior.
Failures can be observed and analyzed at different granularities, from a single transistor to an entire HPC system.
Nodes are the smallest units in HPC systems that have a fully functional computational software stack and can be added or removed from HPC systems with minimum side-effects on the other computing units.
Therefore, the granularity of failure detection in this work is set at the node level.
Heterogeneity of computing nodes, as well as the topology of HPC systems, are other effective factors that influence the overall system performance.
In accordance with node-related effective factors such as physical location, role in the cluster, computational workload, hardware properties, and so forth, various categorizations of computing nodes are possible.
Hereafter, the term \emph{vicinity} is used to refer to nodes that exhibit similar properties such as the ones mentioned above.
The concept of vicinity defines new dimensions of node correlation, beyond the natural temporal and spatial correlations.
\autoref{subsec:vicinity} provides examples and describes the vicinity of the nodes in more detail.

An anomaly is an (observed) unexpected behavior with respect to the expected behavior.
In contrast to failures, unexpected behaviors are not necessarily incorrect behaviors.
Anomaly detection is a well-known general purpose approach for detecting failures in computing systems~\cite{Chandola2012}.
In HPC systems, system log analysis can be used for anomaly detection for the purpose of preventing failures~\cite{park2018, landauer2018, klinkenberg2017, Ghaisvand2016a}.
All HPC systems on the current TOP500~\cite{TOP500} list are Linux-based.
Therefore, they all generate system log (Syslog)~\cite{SYSLOG} messages by default.
The goal of this work is to detect node failures via analyzing fully anonymized system logs using a vicinity-based statistical anomaly detection approach.
The use-case of this study is a production Petascale HPC system, called Taurus\footnote{https://doc.zih.tu-dresden.de/hpc-wiki/bin/view/Compendium/SystemTaurus}.

In addition to technical data, system log entries contain sensitive data about the system and its users.
Therefore, addressing the data privacy~\cite{GDPR} concerns is a fundamental requirement of any mechanism that detects anomalies on production HPC systems via system log analysis.
The anonymization of system log entries, prior to performing Syslog analysis while retaining useful data, addresses the data privacy concerns~\cite{ghiasvandAssessingDataUsefulness25a}.

In this work, the Taurus nodes are first categorized into four different vicinities based on similarities they exhibit in terms of (1)~\emph{hardware architecture}, (2)~\emph{resource allocation}, (3)~\emph{physical location}, and (4)~\emph{time of failures}.
Then, the anomalies are detected within each vicinity.
Subsequently, the effectiveness of performing anomaly detection in each vicinity for the purpose of failure detection is compared and discussed.
To assess the usefulness of the proposed method of anomaly detection on anonymized data, a copy of 8 months of Taurus system logs was anonymized via the PaRS~\cite{PaRS} anonymization mechanism and the anonymized system logs were used as the input data.

The main contributions of this work are: (1)~proposing a node failure detection mechanism via a vicinity-based statistical anomaly detection approach that employs a passive data collection approach, as well as (2)~analyzing the effectiveness of anomaly detection method in various vicinities using the 8 months of Taurus HPC cluster system logs.
In addition, (3)~to the best of our knowledge this is the first work on anomaly detection that is capable of utilizing both original and anonymized Syslog entries with a high degree of equivalence between the analysis results.

The remainder of this work is organized as follows.
The node vicinities, data sources, and anonymization method are introduced in Subsections~\ref{subsec:vicinity}, \ref{sec:data-source}, and \ref{sec:anonymization} respectively.
The method to identify node failures employed in this work is described in \autoref{sec:failures}.
The proposed anomaly detection method is described in detail in \autoref{sec:anomaly}.
The impact of anomaly detection in different node vicinities is analyzed in \autoref{sec:comparison}.
The background and current related works are introduced in \autoref{sec:relatedwork}.
Finally, the work is concluded and future work directions are discussed in \autoref{sec:conclusion}.
\section{Proposed Anomaly Detection Methodology}
\label{sec:methodology}

Taurus is an HPC cluster composed of 2,046 computing nodes.
The computing nodes are organized in 6 islands based on their hardware properties\footnote{Detailed hardware information of Taurus: https://doc.zih.tu-dresden.de/hpc-wiki/bin/view/Compendium/HardwareTaurus}.
\autoref{fig:hardware} provides a schematic illustration of the Taurus node topology.
Each letter represents a single computing node.
Nodes with identical colors are of identical hardware (processing unit) architecture.

\subsection{Vicinities}
\label{subsec:vicinity}
Computing nodes with similar characteristics are considered to be in the \emph{vicinity} of each other.
Node characteristics include any physical, spatial, temporal, or logical properties of the computing nodes.
A group of computing nodes located in the same rack, performing different tasks of the same job, or sharing a common resource (e.g., file system, power supply unit), are all examples of nodes in the vicinity of each other.

The concept of vicinity defines new dimensions of node correlation, beyond the natural temporal and spatial correlations.
Each vicinity can be imagined as a new dimension in which two separated entities (nodes) become correlated.
For example, points $A:(1,10)$ and $B:(4,6)$ in a 2D Cartesian representation are separated by a distance of $4-1=3$ on the $X$ axis and $10-6=4$ on the $Y$ axis, respectively.
Defining the new dimension $Z$, according to a common (but so far unseen) feature of $A$ and $B$ would result in a 3D representation of $A:(1,10,5)$ and $B:(4,6,5)$.
Here '$5$' denotes that common feature.
In the new 3D representation, even though $A$ and $B$ are still separated on $X$ and $Y$, their distance on the dimension Z will be $5-5=0$.
In another word, $A$ and $B$ will be in the vicinity of each other from the $Z$ axis perspective.

In this work, node vicinities are observed from four different perspectives: $(1)~$\emph{hardware architecture}, $(2)~$\emph{resource allocation}, $(3)~$\emph{physical location}, and $(4)~$\emph{time of failure}.
The first perspective denotes a node vicinity according to the node's physical properties, the second perspective emphasizes the node's logical properties, while the third and fourth perspectives denote spatial and temporal properties, respectively.
All other correlations among nodes can be mapped onto these four proposed vicinities, e.g., nodes connected to a single switch can be mapped onto the \emph{physical location} vicinity.
In Subsections~\ref{subsec:hardware},~\ref{subsec:job},~\ref{subsec:location}, and~\ref{subsec:failure} these four vicinities are explained in more detail, based on the Taurus architecture.

The node vicinities are intended to mitigate the major characteristic differences between nodes.
Therefore, in cases that several parameters influence a certain node's characteristic, the most dominant parameter is considered to identify the node's vicinity.
All nodes in island 2 beside their Sandy bridge or Haswell CPUs are equipped with graphical processing units (GPU).
Since the majority of jobs submitted to island 2 mainly utilize GPUs rather than CPUs, GPUs are considered as dominant processing units of these nodes.
Therefore, in this work, island 2 is considered as a homogeneous GPU island, despite the heterogeneity of its nodes' CPUs.

It is important to emphasize that in the context of this work, two nodes in the \emph{vicinity} of each other are not necessarily physically co-located.
In fact, they may even belong to physically separated partitions of the HPC system.

\subsubsection{Hardware Architecture Vicinity}
\label{subsec:hardware}
\begin{figure}
	\begin{center}
		\includegraphics[clip, trim=0.5cm 16cm 1.5cm 0.0cm, width=.45\textwidth]{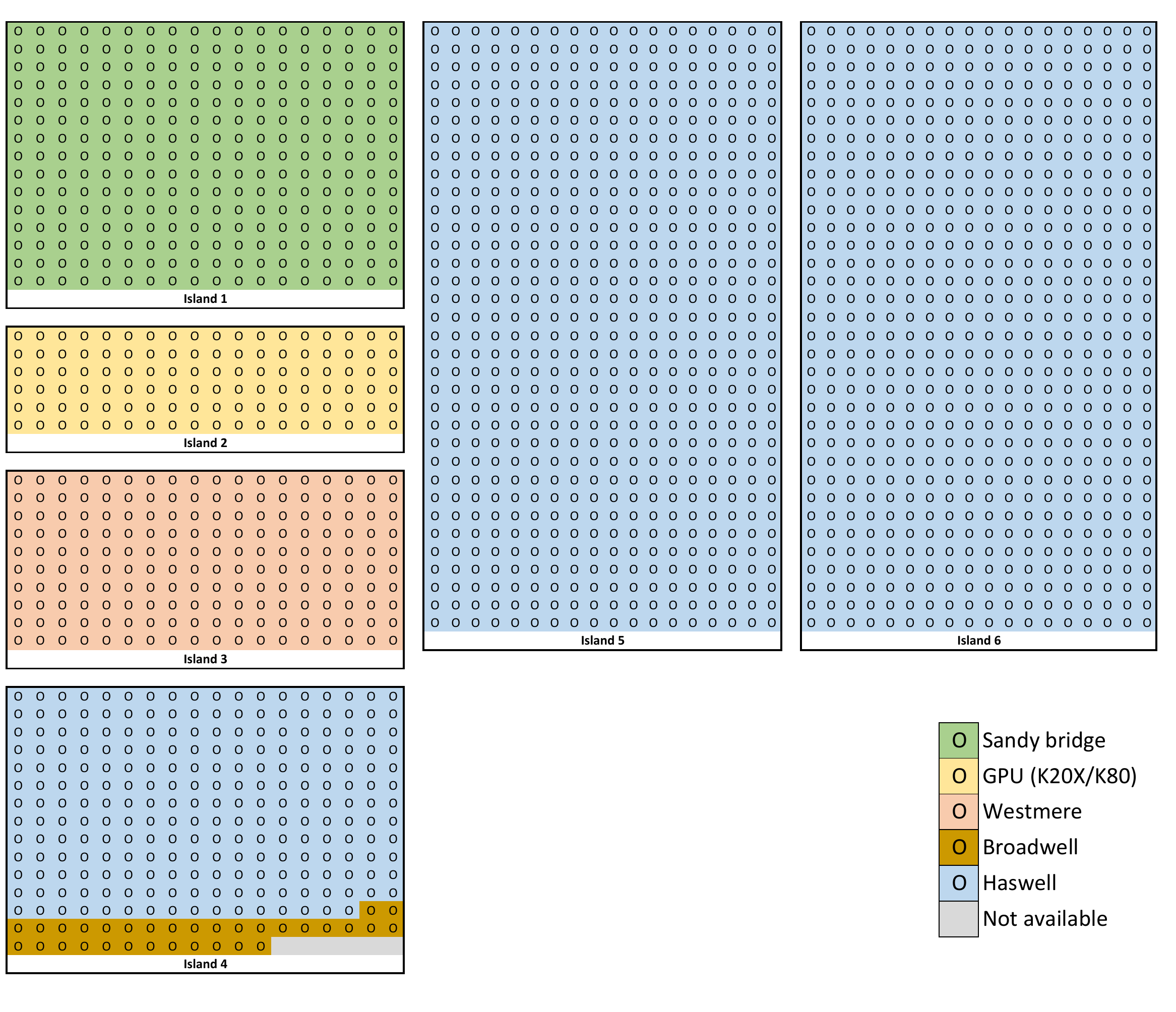}
		\caption{Schematic island and node topology of Taurus. Node colors represent the dominant processing unit type of a node. Thick border lines indicate the 6 islands of Taurus.}
		\label{fig:hardware}
	\end{center}
	\vspace{-0.5cm}
\end{figure}

Computing nodes on Taurus may be of 4 different processors architectures: Intel Haswell, Broadwell, Sandy Bridge, and Westmere.
108 nodes with Sandy Bridge and Haswell processors are also equipped with GPUs (NVIDIA Tesla K20X and K80).
According to their hardware architectures, the 2,046 computing nodes on Taurus can be divided into 5 categories.
The node's dominant processor architecture and the number of nodes in each architecture category are shown in \autoref{tab:hardware}.
A schematic illustration of the Taurus topology, including the type of each node's hardware architecture is provided in \autoref{fig:hardware}.
\textbf{Nodes with identical colors in \autoref{fig:hardware} are in the vicinity of each other from the \emph{hardware architecture} perspective.}

\begin{table}[!t]
	\linespread{0.9}
	\fontsize{6.9pt}{10pt}\selectfont
	\caption{Hardware architecture of Taurus computing nodes}
	\centering
	\addtolength{\tabcolsep}{-4pt}
	\begin{tabular}{l|l|c|c|c|c}
		\textbf{Architecture} & Haswell & Sandy Bridge & Westmere & Broadwell & GPU (K20X/k80) \\
		\hline
		\textbf{Node count} & 1456 & 270 & 180 & 32 & 108 \\
		\textbf{Island} & 4, 5, 6 & 1 & 3 & 4 & 2 \\
	\end{tabular}
\label{tab:hardware}
\end{table}

\subsubsection{Resource Allocation Vicinity}
\label{subsec:job}
Taurus employs Slurm~\cite{slurm} for job scheduling.
The submitted jobs are allocated resources based on direct requests of the users, system policies, and the status of available resources.
\textbf{All nodes that execute tasks of the same job are in the vicinity of each other from the \emph{resource allocation} perspective.}
In contrast to the static nature of the hardware architecture vicinity perspective, the resource allocation vicinity is fully dynamic and may change frequently as the running jobs are completed and new jobs are submitted to the cluster.

\subsubsection{Physical Location Vicinity}
\label{subsec:location}

Various granularities can be used to express the physical location of a node in Taurus, e.g., chassis, rack, or island.
Since the power, temperature, and connectivity of all nodes located in a single rack are controlled together, this work considers racks as the physical location granularity.
Each row of an island shown in \autoref{fig:hardware} represents one rack of nodes.
\textbf{All nodes located in the same rack are in the vicinity of each other from the \emph{physical location} perspective.}

\subsubsection{Time of Failure Vicinity}
\label{subsec:failure}
Often failure is a consequence of various node-level events on and of properties of several nodes.
However, a failure in itself is observable on a particular node at a specific moment.
Therefore, the time of failure is considered as a temporal property of that particular node even though, several nodes may fail due to the same reason.
\textbf{From this perspective, all nodes that experience a failure within the same predefined time interval, fall into the same vicinity category.}
In this work, the time of failure interval considered is 10 minutes.
The 10-minute time interval is chosen according to the results of the previous study on Taurus failure correlations~\cite{Ghaisvand2016a}.
That study revealed that the majority of failures correlated on Taurus occurred within 10 minutes of each other.
Therefore, failures that occur across the entire system within 10 minutes of each other are assumed to be in the same temporal vicinity.
Thus, the nodes on which such failures occur are in the vicinity of each other from the \emph{time of failure} perspective.

\subsection{Data Sources}
\label{sec:data-source}

System logs represent the main data source in this work.
Syslogs of Taurus were collected for a period of one year from January to December 2017.
Syslog daemons on each node collected the system log entries and forwarded them to a central log collector.
To maintain the neutrality of the results and to provide a general approach, the pre-configured Syslog daemons were used without any changes, except for being configured to forward Syslog entries to the central log collector.

In addition to Syslog entries, three other data sources were considered to improve the accuracy of the failure identification method: \emph{outage database}, \emph{service notifications}, and \emph{job status reports}.
The outage database reports all system outages from the users' perspective, the service notifications notify users regarding future scheduled maintenances and system-wide outages, and the job status reports indicate the final status of a job after its execution.
\autoref{tab:data-sources} provides an overview of the four data sources used in this work.

\begin{table}[!t]
	\linespread{0.9}
	\fontsize{6.9pt}{10pt}\selectfont
	\caption{Data sources}
	\centering
	\begin{tabular}{l|l|l|l}
		\textbf{Data source} & \textbf{Data collection} & \textbf{Information} & \textbf{Granularity} \\
		\hline
		System logs & Automatic & Software and hardware & Component \\
		Outage database & Semi-automatic & Service availability & Entire system \\
		Service notifications & Manual & Service availability & Entire system \\
		Job status reports & Automatic & Job completion status & Node \\
		
	\end{tabular}
	\label{tab:data-sources}
\end{table}

\begin{figure}
	\begin{center}
		\includegraphics[trim= 0 0 20 60, clip, width=.45\textwidth]{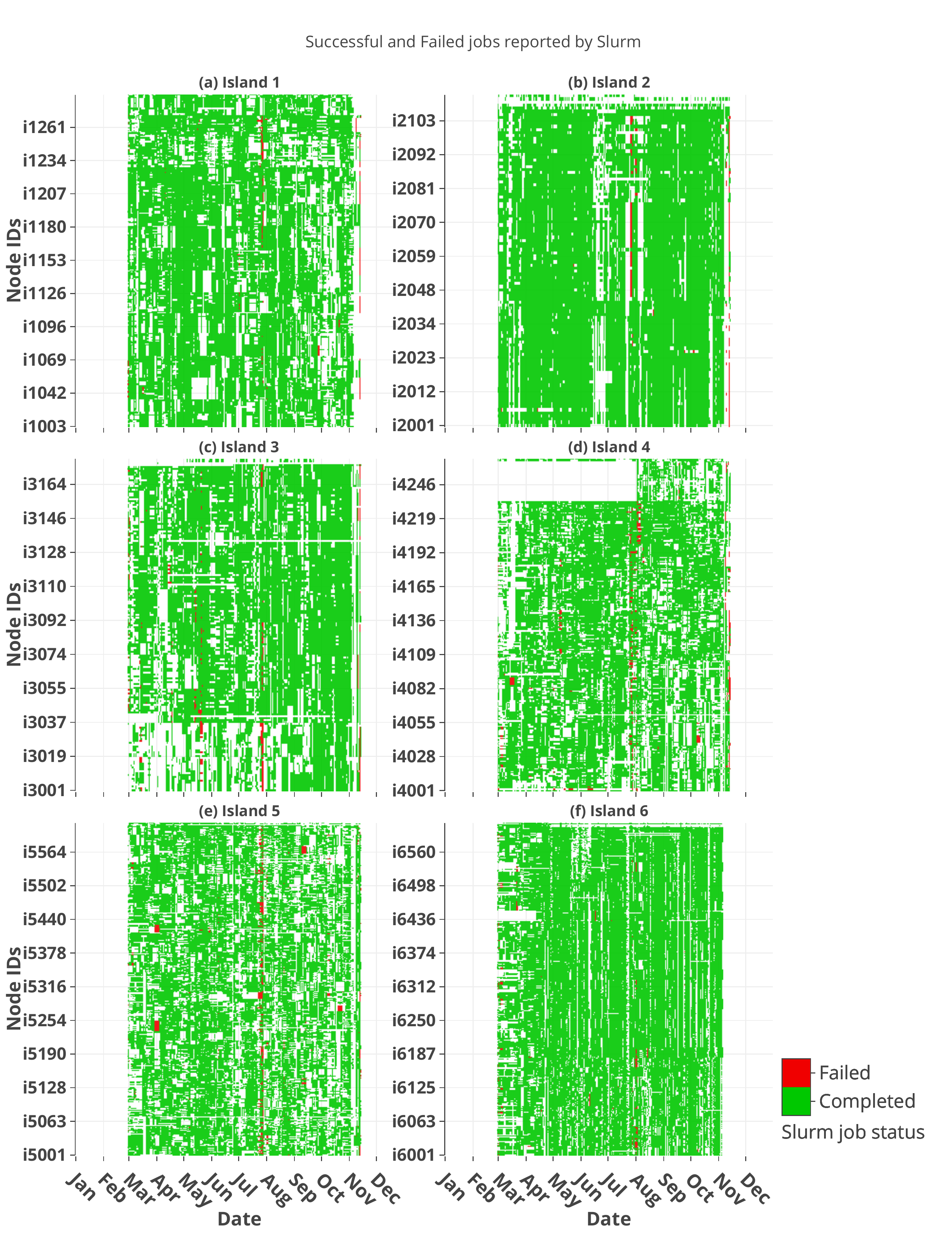}
		\caption{Slurm job status report on all islands of Taurus for the year 2017. Intervals of unavailability of jobs reports do not necessarily specify node outages. The red dots indicate jobs reported as "failed" concurrent to a node failure.}
		\label{fig:slurm-failure}
	\end{center}
\end{figure}

There are certain data gaps in the four data sources used in this work.
The gaps are mainly incurred due to the interruption of the data collection mechanism.
The Syslog entries cover the full period of one year from January to December 2017.
The job status reports generated by Slurm showed in \autoref{fig:slurm-failure} covers the period of 28-02-2017 to 14-11-2017.
The service notifications and the outage database, provide information from a higher perspective and are available for the whole period of one year from January to December 2017.

Given these available data, the focus of this work is on the period of 01-03-2017 to 31-10-2017.
The existence of certain gaps in data sources is reportedly a common challenge in similar studies \cite{Sayed2013}.

\subsection{Anonymization of System Logs}
\label{sec:anonymization}

System log entries contain sensitive data about the system and its users.
Therefore, addressing data privacy concerns is a fundamental requirement of any mechanism that performs system log analysis.
The anonymization of system log entries, prior to performing Syslog analysis, addresses the data privacy concerns.

\emph{PaRS}~\cite{PaRS} is an anonymization mechanism that provides full anonymization through hashing the message part of Syslog entries.
PaRS substitutes all variables and sensitive terms in each log entry with constant values and maps each substituted entry to a hash key.
Through anonymization, the partial semantic of log entries required for anomaly detection via Syslog analysis is preserved.
The rightmost column in \autoref{tab:anonymization} contains the anonymized Syslog entries which correspond to the raw log entries shown in the middle column (prior to anonymization).
As shown in \autoref{tab:anonymization}, an identical hash key is generated for entries with similar raw log semantics.

\begin{table}[!t]
	\fontsize{6.3pt}{10pt}\selectfont
	\caption{Sample syslog entries in their original and anonymized form}
	\centering

	\begin{tabular}{l|l|l|l}
		\#&
		\textbf{Original Syslog entry} &
		\textbf{Anonymized} \\
		\hline
		1 & \verb|(siavash) CMD (/usr/bin/check >/dev/null 2>&1)| & \cellcolor{blue!15}66dc2742 \\
		2 & \verb|(florina) CMD (/usr/lib32/lm/lm1 1 1)| & \cellcolor{blue!15}66dc2742 \\
		3 & \verb|(siavash) CMD (run-parts /etc/cron.hourly)| & \cellcolor{blue!15}66dc2742 \\
		4 & \verb|starting 0anacron| & \cellcolor{red!15}dd740712 \\
		5 & \verb|Anacron started on 2018-01-30| & \cellcolor{green!15}e5a59462 \\
		6 & \verb|Jobs will be executed sequentially| & \cellcolor{yellow!15}f1e7eac3\\
		7 & \verb|Normal exit (0 jobs run)| & \cellcolor{brown!15}eac7924f \\
		8 & \verb|finished 0anacron| & \cellcolor{pink!15}a5803a8a \\
		9 & \verb|(siavash) CMD (/usr/lib32/lm/lm1 1 1)| & \cellcolor{blue!15}66dc2742 \\
		10& \verb|(root) CMD (/usr/lib32/cl/cl2 1 1)| & \cellcolor{blue!15}66dc2742 \\	
	\end{tabular}

	\label{tab:anonymization}
	\vspace{-0.5cm}
\end{table}

\subsection{Taurus Node Failures}
\label{sec:failures}

To assess the proposed method's functionality, all Taurus node failures must be known.
Due to various technical reasons, the complete list of all node failures on Taurus for the period of this work is not available.
This step is aimed to provide a complete list of Taurus node failures as the ground truth for further analysis and comparisons in \autoref{sec:comparison}.

\begin{figure}
	\begin{center}
		\includegraphics[trim= 55 40 55 40, clip, width=.45\textwidth]{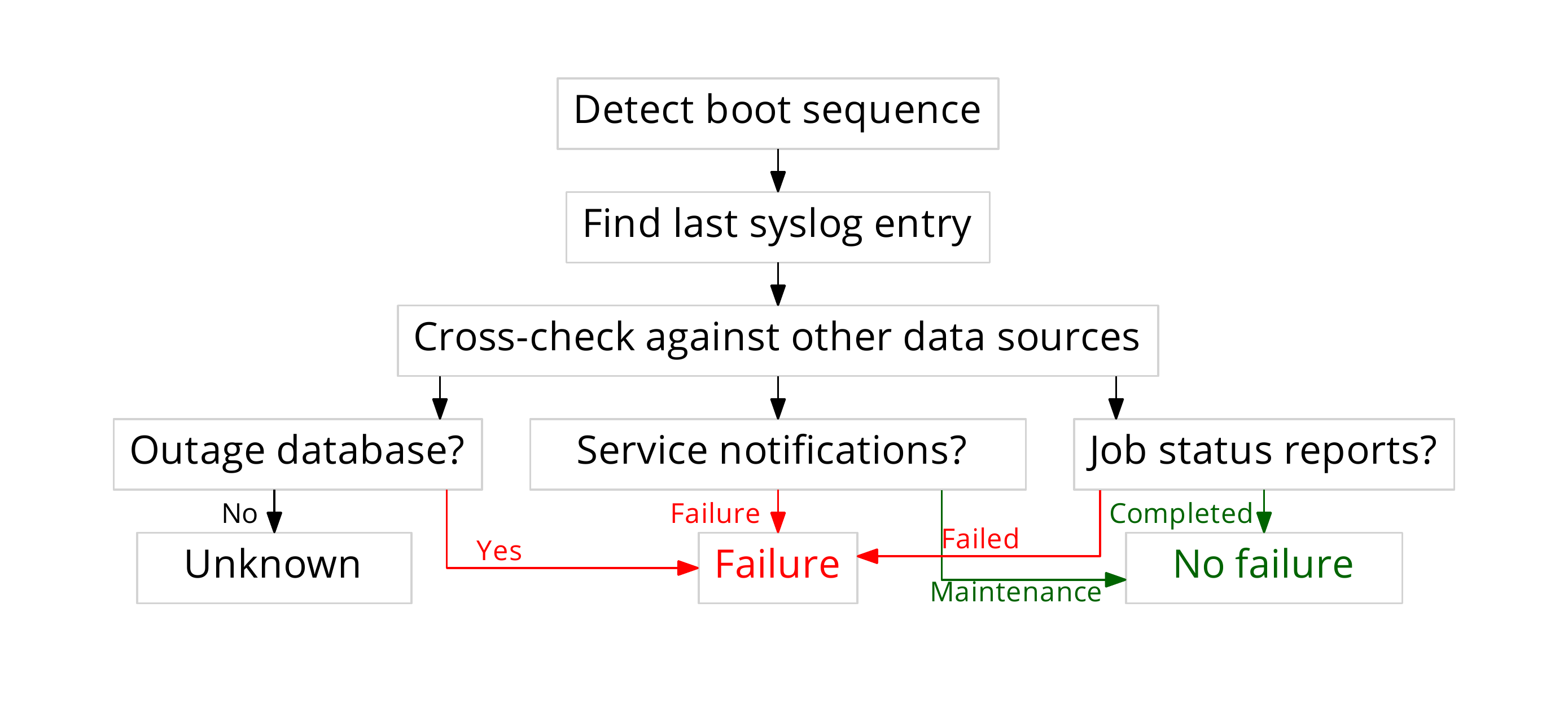}
		\caption{Failure detection Workflow via system logs. In case of having both contradicting results at once, the incident will not be considered as a failure.}
		\label{fig:workflow}
	\end{center}
	\vspace{-0.5cm}
\end{figure}

Node failures in computing systems can be divided into two main categories according to their root causes:
(1)~Failures that occur during the normal operation of the HPC system caused by internal factors, such as software and hardware errors or race conditions (i.e., regular failures).
(2)~Failures that occur due to external factors, such as power outages and human errors.
Analyzing the impact of the external causes of node failures requires additional information regarding external factors that are not available, e.g., detailed information about the behavior of the HPC system power supply.
Therefore, the focus in this work is on the first group of node failures namely \emph{regular failures}, typically caused by internal factors.
The next step to identify such failures is to detect node outages and to distinguish regular failures from those which may occur as a result of external factors, such as maintenance, human errors, and others.
The failure detection workflow in Taurus is shown in \autoref{fig:workflow}.

Computing nodes of Taurus generate and send Syslog entries to a central log collector which stores them for future analysis.
This is a passive log collection mechanism chosen due to imposing no additional overhead, and to be applicable to other HPC systems.
However, the failure identification process becomes more challenging in comparison to the use of active log collection mechanisms.
Due to using the passive log collection mechanism, a node outage can be confidently detected only when a direct indication in the form of a log entry is generated by the failing node and correctly received and stored by the central log collector, e.g.," Kernel panic - not syncing: Fatal Exception."
However, in many cases, a node outage leaves no direct indication in system logs.
A workaround is to assume the absence of log entries for more than a certain period of time as an indication of a potential outage.
Nonetheless, this assumption is not accurate.
For various reasons such as CPU overloading or network congestion, the flow of system log entries from the computing nodes to the central log collector may be interrupted or delayed, which could be assumed as an outage, even though the computing nodes are functional.
Also, in many cases immediately after the occurrence of an outage, the protection mechanisms recover the node.
In both latter scenarios (temporary interruption in data flow and automatic node recovery), an active node probing approach may also fail to correctly detect all node outages.

Analyzing Taurus system logs revealed that during a healthy boot all nodes leave similar footprints in Syslog entries.
When a node fails to generate the expected footprint at boot time, it is an indication of a faulty boot process and thus the node will be either automatically rebooted or it will fail shortly thereafter.
The higher frequency of log generation at boot time in comparison with the normal operation is another indicator of a boot event, which can be used to identify even a problematic boot process.

\begin{figure}
	\begin{center}
		\includegraphics[trim= 0 0 20 60, clip, width=.45\textwidth]{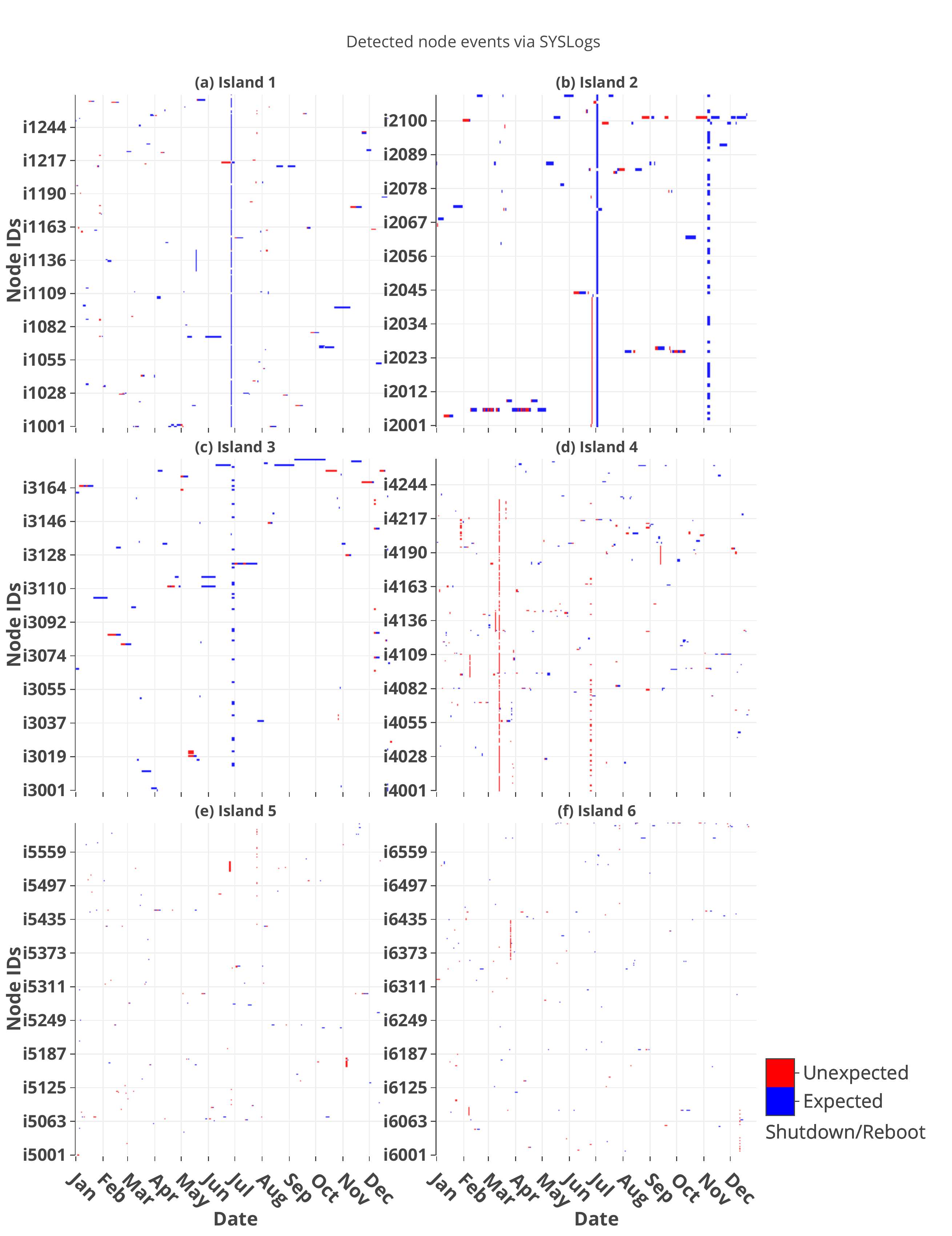}
		\caption{Node outages detected via Syslog analysis. Unexpected events (shown in red) indicate the absence of information in system logs which may be a sign of node crashes. Expected events indicate node outages that are planned due to maintenance or intentionally caused by the system protection mechanisms in place.}
		\label{fig:syslog-events}
	\end{center}
	\vspace{-0.2cm}
\end{figure}

The proposed node outage detection method in this work first detects the node boot events.
Afterward, Syslog entries are backtracked until the last Syslog entry before the boot event is found.
The timestamp of the last Syslog entry prior to the boot event is considered as the point of the outage.
All node outages will be identified using the proposed method.
The only exception is when a node fails and has no further successful boot process.
In such cases, comparing the timestamp of the last available Syslog entry with the current time (i.e., $31$-$10$-$2017$ $23$:$59$:$59$) reveals the missing node outages.
\autoref{fig:syslog-events} illustrates all detected node outages on Taurus over the course of 2017.

The detected outages are compared against the information from other available data sources (mentioned in \autoref{tab:data-sources}).
When a node outage occurs outside of the scheduled maintenance period and no job could be accomplished on that particular node at the time of the detected outage, the detected outage represents a \emph{regular failure}.
As \autoref{fig:slurm-failure} illustrates, it is common that certain jobs on a specific node fail, although other jobs on the same node are accomplished simultaneously.
Also, when a node outage is recorded in the outages database that monitors the availability of the HPC system from the users' perspective, it is considered as a \emph{regular failure}.

\subsection{Anomaly Detection}
\label{sec:anomaly}
\begin{figure}
	\begin{center}
		\includegraphics[trim= 50 540 140 90, clip, width=.45\textwidth]{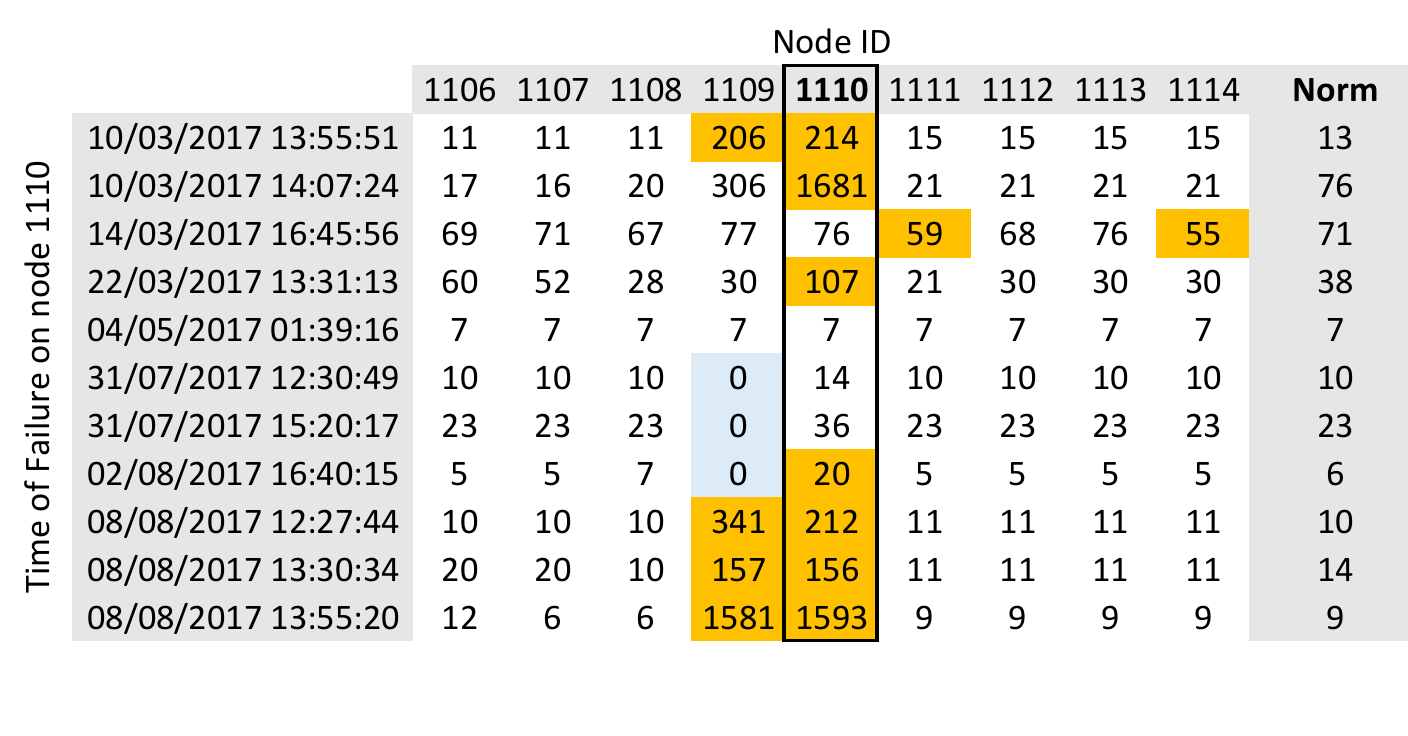}
		\caption{Detection of failures (shown in orange) of node 1110 via the proposed failure detection mechanism. Cells colored in light blue indicate non-responsive nodes.}
		\label{fig:behavior_sample}
	\end{center}
	\vspace{-0.5cm}
\end{figure}

The common behavior of most nodes within a node vicinity is considered as \emph{normal} behavior in that vicinity.
A node's behavior is defined as the \textbf{S}yslog \textbf{G}eneration frequency of the node (hereafter SG).
The SG parameter is dynamically calculated based on the number of Syslog entries received from a computing node during a fixed time window (e.g. 30 minutes) prior to the current (observation) moment.
The SG parameter of each node is compared against the SG of other nodes in the same vicinity.
Based on these comparisons, the normal value of the SG parameter of certain nodes at a given moment in time is calculated.
Once the deviation of a node's SG parameter from the normal behavior exceeds a certain threshold, the node's behavior is considered \emph{abnormal}.
The deviation threshold is dynamically calculated within each vicinity\footnote{A sample code written in python to demonstrate the calculation of dynamic thresholds via k-means is available: ghiasvand.net/u/param}.
To calculate the deviation threshold, all nodes within a vicinity (i.e. one row of \autoref{fig:behavior_sample}) are partitioned into 2 clusters based on their SG parameter via a clustering method such as K-Means.
The deviation threshold is the relative density of resulting clusters which is calculated as the sum of squared distances of samples to their closest cluster center\footnote{Also known as \texttt{within cluster sum}.}.

\begin{figure}
	\begin{center}
		\includegraphics[trim= 50 540 140 90, clip, width=.45\textwidth]{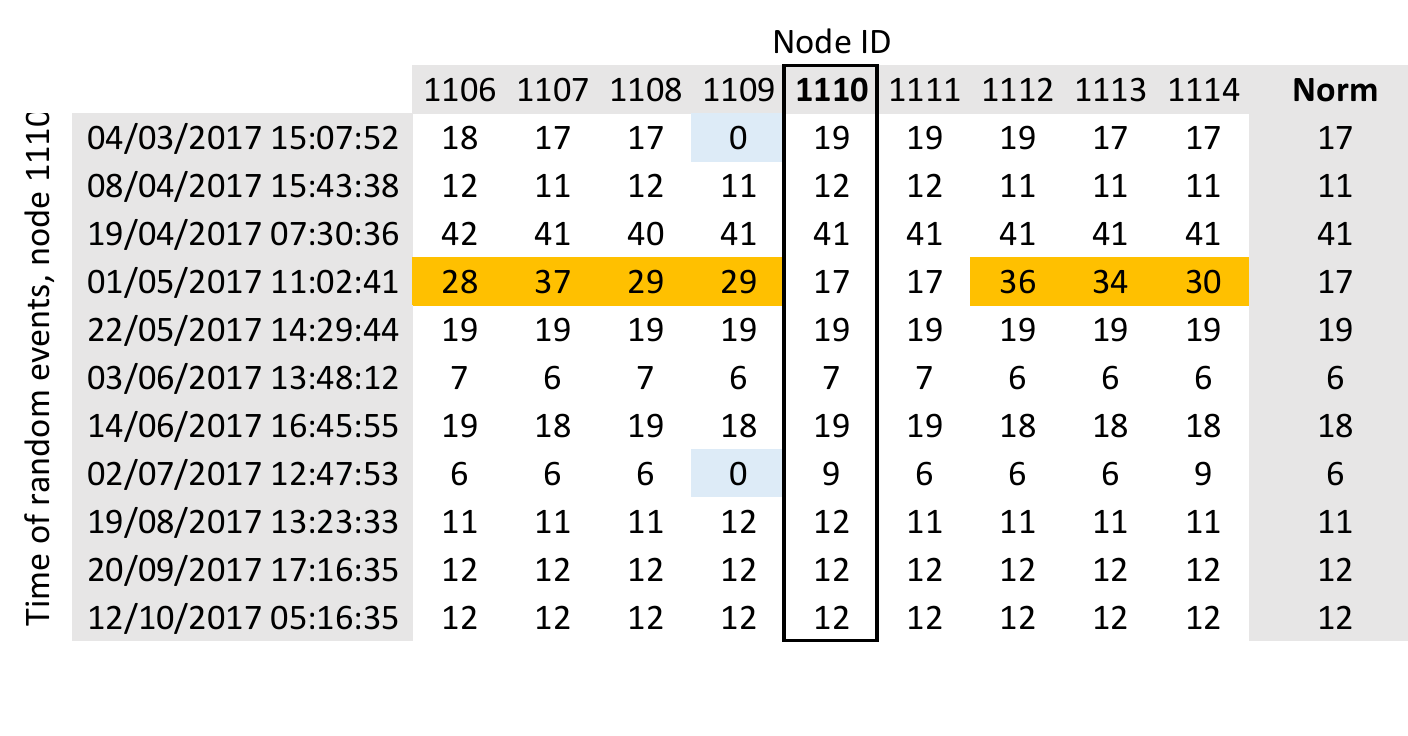}
		\caption{Application of the proposed failure detection mechanism during normal behavior of node 1110. Detected failures are shown in orange. Cells colored in light blue indicate non-responsive nodes.}
		\label{fig:behavior_random}
	\end{center}
	\vspace{-0.5cm}
\end{figure}

\autoref{fig:behavior_sample} illustrates the behavior of node 1110 and  8 other neighboring nodes (physically located next to each other) prior to a point in time when 11 failures occurred on node 1110 in the year 2017.
\autoref{fig:behavior_random} on the other hand, illustrates the behavior of the same nodes prior to 11 random points of time in which node 1110 functioned normally.
In \autoref{fig:behavior_sample} and \autoref{fig:behavior_random}, the timestamp at the beginning of each row represents the observation moment in the year 2017.
The value of each cell represents the SG parameter of the respective node (column's header) within a time interval of 30 minutes prior to the observation moment.
Cells with abnormal behavior are shown in orange.
The cell coloring in each row is relative to the value of other cells in that particular vicinity (row).
According to \autoref{fig:behavior_sample}, node 1110 experienced 11 failures in 2017.
For 7 out of the 11 failures illustrated in \autoref{fig:behavior_sample}, the deviation of the SG parameter correctly identifies the abnormal behavior of node 1110.

In the example provided in \autoref{fig:behavior_sample}, the SG parameters were obtained for the nodes \emph{physical location} vicinity.
The same comparisons were made within other node vicinities.
In \autoref{sec:comparison} the effectiveness of the proposed anomaly detection method in each node vicinity is discussed.

\section{Impact of Vicinities on Anomaly Detection}
\label{sec:comparison}
The proposed anomaly detection method was applied to 8 months of Taurus system logs.
Regular failures were detected based on the node vicinities introduced in \autoref{subsec:vicinity}.
The proposed method is applicable to \emph{hardware architecture}, \emph{resource allocation}, and \emph{physical location} vicinities for the purpose of online anomaly detection.
However, the \emph{time of failure} vicinity can only be used as an offline approach to analyze the nodes' behavior after each failure occurrence.

Results of the proposed method's application on Taurus system logs were compared against the set of regular failures detected in \autoref{sec:failures}.
The following subsections describe the impact of performing the proposed anomaly detection method in each vicinity in more detail.

\subsection{Impact of Hardware Architecture Vicinity}
\label{subsec:impact-hardware}
Taurus nodes are located in 6 islands.
As shown in \autoref{fig:hardware}, island 4 hosts nodes with different processor types, while islands 1, 2, 3, 5, and 6 are homogeneous.
Although the nodes' hardware architecture influences the job allocation, as \autoref{fig:slurm-failure} illustrates there is no noticeable difference among job allocation patterns on various Taurus islands.
However, as shown in \autoref{fig:syslog-events}, with the exception of islands 5 and 6 -which comprise of identical processor types- the node outages have different distribution patterns on each island.

\begin{figure}
	\begin{center}
		\includegraphics[trim= 13 0 60 60, clip, width=.45\textwidth]{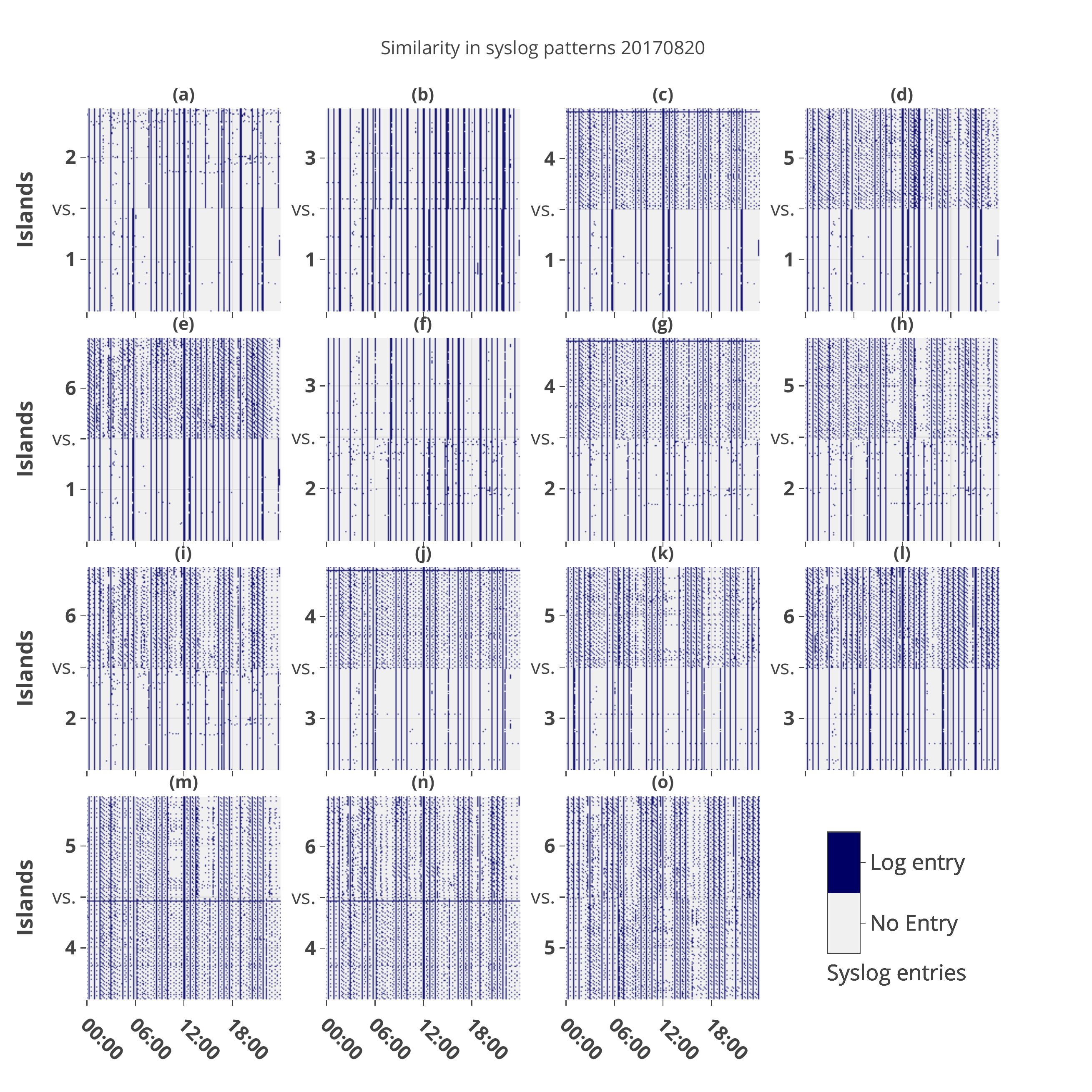}
		\caption{Syslog generation patterns of different islands. Each sub-diagram is vertically divided into two sections. Each section illustrates the Syslog generation pattern of 100 nodes of the respective island during 24 hours. e.g., sub-diagram (e) illustrates the Syslog generation pattern of island 1 (bottom) versus island 6 (top).}
		\label{fig:syslog-pattern-comparision}
	\end{center}
	\vspace{-0.5cm}
\end{figure}

\autoref{fig:syslog-pattern-comparision} illustrates a one-to-one comparison of Syslog generation patterns in all 6 islands of Taurus.
This figure visualizes the temporal and spatial patterns among more than 46K, 82K, 45K, 968K, 940K, and 1M Syslog entries generated by islands 1 to 6, respectively.
Islands 5 and 6 present an almost identical pattern, which is also very similar to island~4.
In contrast, the system log generation pattern on each of the islands 1, 2, and 3 has a completely different pattern.

The comparison shown in \autoref{fig:syslog-pattern-comparision} indicates that the processor architecture has a direct impact on node behavior.
Therefore, the behavior of nodes in island 1 (Sandy Bridge) should not be predicted based on the behavior of nodes in island 3 (Westmere), while a similar behavior is expected from nodes in island 5 (Haswell) and island 6 (Haswell).

No additional patterns were detected when conducting a similar analysis based on the amount of node's physical memory.
The use of the proposed anomaly detection method on nodes with different hardware architecture proved to be virtually impossible.
Detecting anomalies within the hardware architecture vicinity on Taurus also revealed several false positives.

It is intended to improve the accuracy of the proposed statistical anomaly detection method by identifying the most relevant vicinity perspective among the four vicinities considered in this work.
The proposed method detects anomalies by analyzing fully anonymized system logs.
To the best of our knowledge, there is no similar approach for detecting anomalies using fully anonymized system logs.
Therefore, a quantitative comparison cannot be conducted.
However, \autoref{tab:discussion} shows a qualitative comparison of the proposed method's accuracy inside and outside of each vicinity.

\subsection{Impact of Resource Allocation Vicinity}
\label{subsec:impact-resource}
\autoref{fig:slurm-failure} illustrates jobs that failed due to node failures, as reported by Slurm.
Several jobs which were allocated on multiple nodes were terminated due to node failures during the 8-month period considered in this work. 
However, except for one incident in which a misbehaving job was causing the node failures, no direct correlation between the node failures and resource allocation was found.
Therefore, among a group of computing nodes executing various tasks of the same job, as long as the running job is not causing a node failure itself, the probability of a second node failure occurrence is not higher than on the rest of the cluster.

It is important to emphasize that the globally shared resources are not included in this analysis and conclusion.
The abnormal behavior of globally shared resources such as the distributed file system may cause node failures itself.
Globally shared resources are, however, unique entities in the cluster.
Thus, directly including them in anomaly detection will not bring any additional benefit.
Globally shared resources are, in fact, indirectly considered in anomaly detection as correlations between nodes.

The use of the proposed anomaly detection method on resource allocation vicinity did not improve the accuracy of anomaly detection.
However, the low number of failed jobs allocated on multiple nodes prevents drawing a robust conclusion.

\subsection{Impact of Physical Location Vicinity}
\label{subsec:impact-location}
Computing nodes are located inside HPC systems according to their function and architecture.
Therefore, the physical location of the nodes is not an independent factor.
The physical location of each computing node on Taurus can uniquely be addressed by its island, rack and chassis number.
Analyzing the distribution of failures on Taurus as shown in Figures~\ref{fig:slurm-failure} and \ref{fig:syslog-events}, reveals a strong correlation among failures in each rack.
Therefore, in comparison with the rest of the cluster, it is more likely to observe the second failure in a rack that already experienced a node failure.

\begin{figure}
	\begin{center}
		\includegraphics[clip, trim=70 130 70 110, width=.45\textwidth]{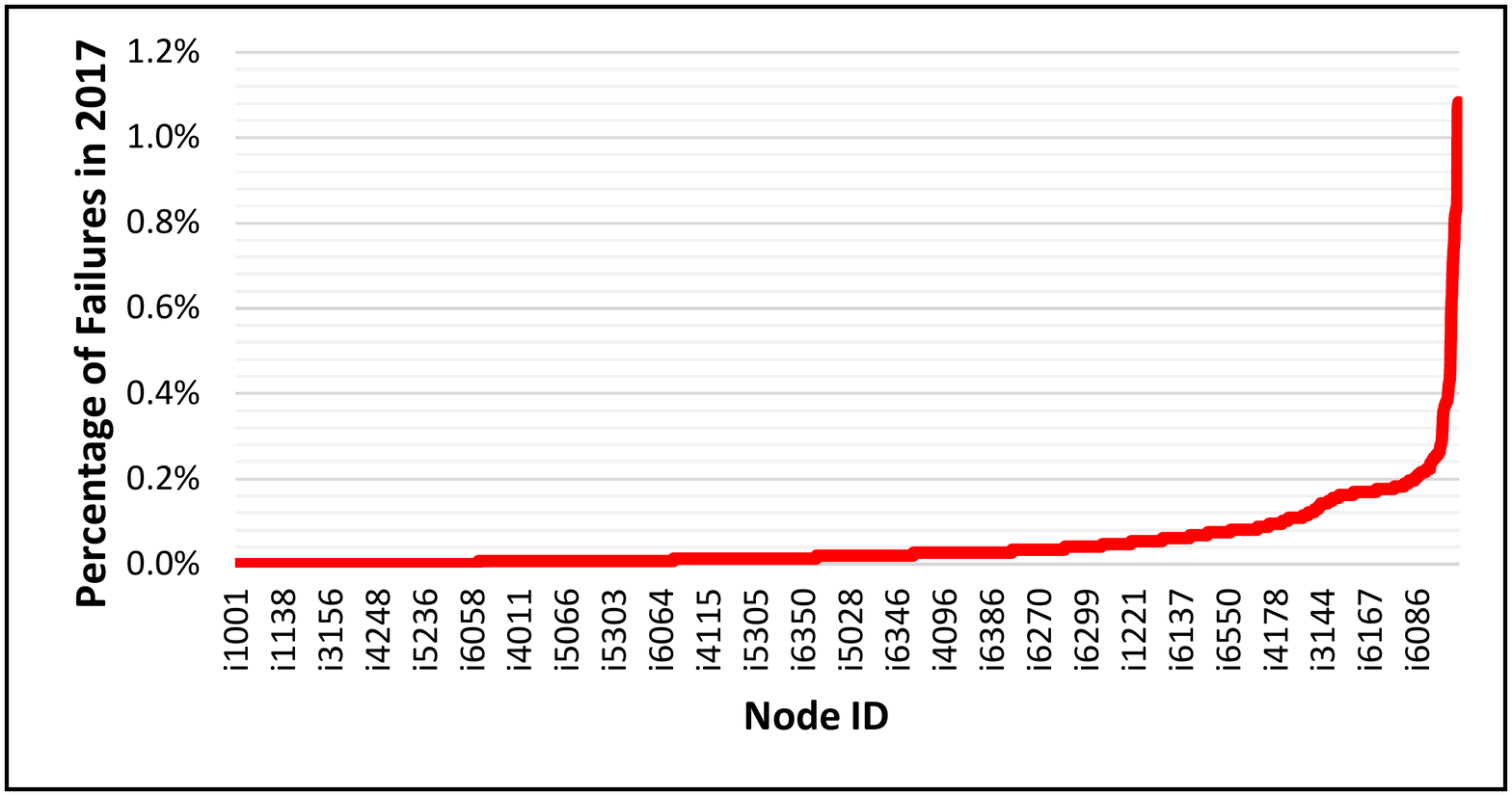}
		\caption{Taurus node failures in the year 2017, nodes are sorted increasingly by the number of failures. The exponential increase in the number of failures on certain nodes indicates a higher probability of subsequent failures.}
		\label{fig:job-failure}
	\end{center}
\end{figure}

Out of the 2,046 computing nodes on Taurus, about $20\%$ of nodes never reported a node failure, and $60\%$ of computing nodes experienced 10 or fewer node failures during the time period considered in this work. 
The remaining $20\%$ of computing nodes are responsible for more than $70\%$ of all node failures.
As shown in \autoref{fig:job-failure}, the probability of a future node failure is in proportion to the number of previous node failures.

Application of the proposed anomaly detection method on the physical location vicinity provided the most accurate results among the three vicinities of hardware architecture, resource allocation, and physical location.
The higher accuracy and the static nature of physical location vicinity make it a good candidate for the application of the proposed anomaly detection method.

\subsection{Impact of Time of Failure Vicinity}
\label{subsec:impact-time}
The \emph{time of failure} vicinity includes all computing nodes which experience a node failure within a predefined time period (e.g., 10 minutes) regardless of their physical location.
Therefore, it is required to regularly update the set of nodes in this vicinity and also unlike the others, it can be used only for offline behavior analysis after node failure.
However, the results of offline behavior analysis can be used in online anomaly detection. 

\begin{figure}
	\begin{center}
		\includegraphics[width=.45\textwidth]{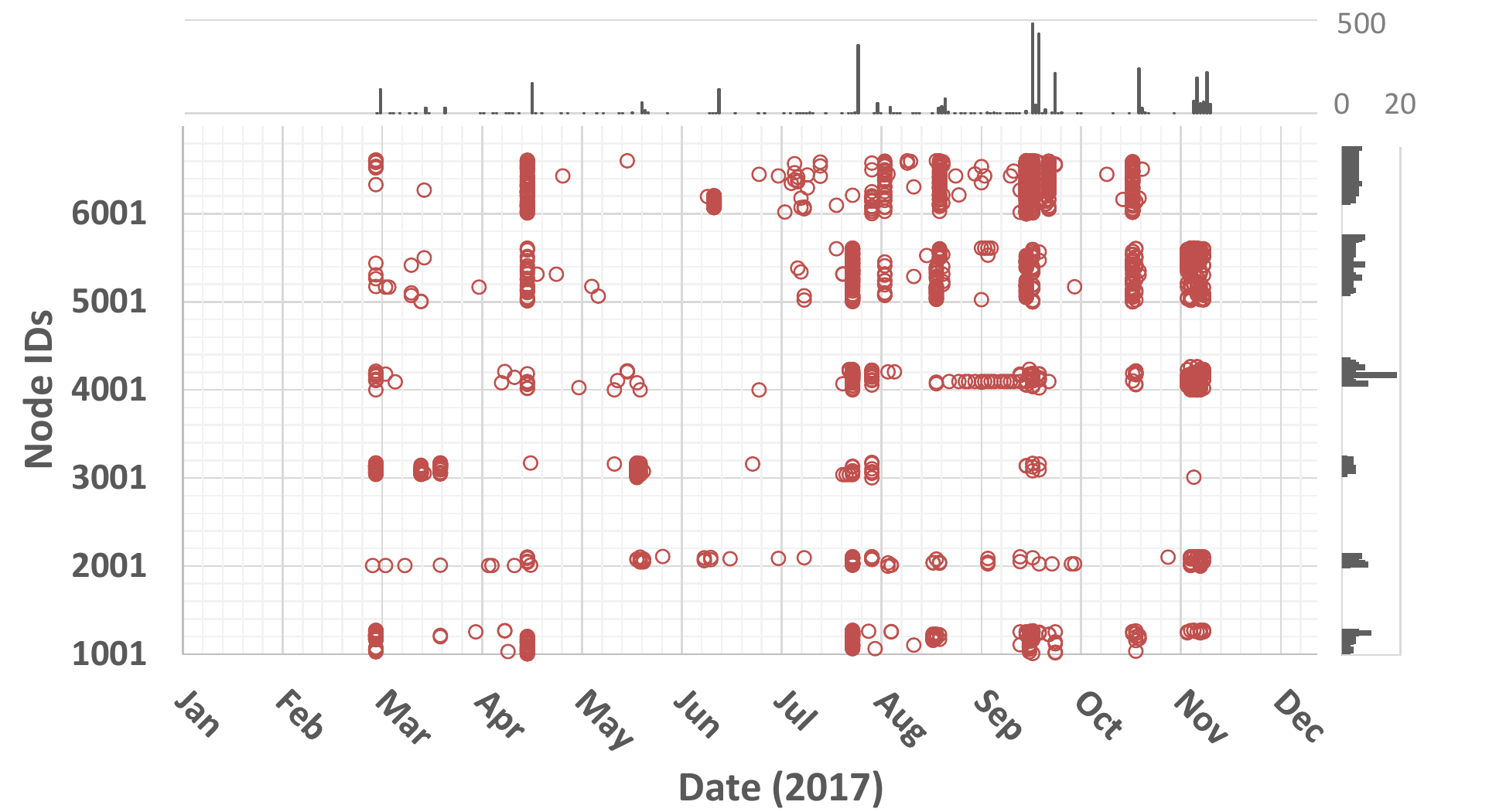}
		\caption{Jobs reported as "failed" on Taurus in 2017 at the time of node failures.}
		\label{fig:job-failure-time}
	\end{center}
	\vspace{-0.5cm}
\end{figure}

\autoref{fig:job-failure-time} illustrates job failures on Taurus.
Each red circle represents a failure occurred on a node on a specific date.
The horizontal concentration of red circles indicates several failures on a node, while the vertical concentration of circles represents simultaneous failures on several nodes.
In several cases, a temporal correlation among node failures is observable.
However, in most cases, the temporal correlation is supported by an even stronger spatial correlation (horizontal concentration of points in \autoref{fig:syslog-events} and \autoref{fig:job-failure-time}).

Application of the proposed anomaly detection method on \emph{time of failure} vicinity provided different results according to the cause of failures.
When failures were caused by misbehaving global shared resources, such as the distributed file system, the results were accurate and reliable.
However, for other causes of failures, such as misuse of local resources or sudden network interruptions, a high number of false positives hindered the successful anomaly detection.

\subsection{Results and Discussions}
\label{discussions}

\autoref{tab:discussion} summarizes the outcome of Subsections~\ref{subsec:impact-hardware}, \ref{subsec:impact-resource}, \ref{subsec:impact-location}, and \ref{subsec:impact-time} in which the efficiency of anomaly detection was studied in various vicinities.
According to these preliminary results, the impact of failure detection via the proposed method inside \emph{resource allocation} and \emph{time of failure} vicinities of Taurus is negligible and thus, should be avoided.
Failure detection inside the \emph{Physical location} vicinity, on the other hand, has a high impact on the accuracy of the final results.
It also became evident that the application of the proposed method outside the \emph{hardware architecture} vicinity of nodes can significantly degrade the accuracy of the failure detection mechanism.

\begin{table}[!t]
	\fontsize{6.9pt}{10pt}\selectfont
	\caption{Accuracy of the proposed failure detection method in different vicinities}
	\centering

	\begin{tabular}{l|l|l|l|l}
		  \multicolumn{1}{p{1.1cm}|}{\centering \textbf{Anomaly} \\ \textbf{detection}}
		& \multicolumn{1}{p{1.1cm}|}{\centering \textbf{Hardware} \\ \textbf{architecture}}
		& \multicolumn{1}{p{1cm}|}{\centering \textbf{Resource} \\ \textbf{allocation}}
		& \multicolumn{1}{p{1cm}|}{\centering \textbf{Physical} \\ \textbf{location}}
		& \multicolumn{1}{p{1cm}}{\centering \textbf{Time of} \\ \textbf{failure}} \\
		\hline
		Inside vicinity & Fair & Low & High & Fair (certain failures) \\
		\hline
		Outside vicinity & No & Low & Low & No \\

	\end{tabular}

	\label{tab:discussion}
\end{table}

Therefore, based on the preliminary results shown in \autoref{tab:discussion} the proposed method was applied to the \emph{physical location} vicinities within each \emph{hardware architecture} vicinity.
With the exception of one rack in island 4 as shown in \autoref{fig:hardware}, on Taurus this vicinity is practically a single rack with homogeneous computing nodes.

\begin{table}[!t]
	\linespread{0.9}
	\fontsize{6.9pt}{10pt}\selectfont
	\caption{Results of applying the proposed failure detection method on Taurus Syslogs entries}
	\centering
	\begin{tabular}{l|l|l}
		 \textbf{Data format} & \textbf{Precision} & \textbf{Recall} \\
		\hline
		Syslog & $62\%$ & $89\%$ \\
		\hline
		Anonymized Syslog & $62\%$ & $89\%$ \\
		\hline
		Filtered Syslog & $87\%$ & $83\%$ \\			
		\hline
		Filtered anonymized Syslog & $81\%$ & $75\%$ \\	
	\end{tabular}
	\label{tab:results}
\end{table}

\autoref{tab:results} summarizes the final results of applying the proposed failure detection mechanism on Taurus system logs.
Using the original Syslog entries as the input data, with a precision of $62\%$ the majority of failures were detected.
Since the proposed method only considers the frequency of log entries rather than the content of each log entry, the similar output could be achieved by using the anonymized Syslog entries without endangering the users' privacy.
The PaRS anonymization method, used in this work, preserves the similarity of Syslog entries such that the frequency of each type of entries can be precisely calculated.
Filtering out frequent log entries from the input data further improves the precision, with a $6\%$ penalty on recall.
Filtering frequent entries from anonymized system logs also improves the precision of the failure detection mechanism, although the recall factor is reduced to $75\%$.
The difference between the results of filtered Syslog entries and filtered anonymized Syslog entries lies in the filtering approach.
Since the content of anonymized system logs cannot be read, frequent log entries are detected based on their hash key and time-stamp.
Therefore, some log entries have been mistakenly filtered out.
Filtering frequent log entries before data anonymization requires further analysis of original system logs and may endanger user privacy.

The majority of the undetected anomalies (false negatives) are related to sudden failures that do not leave large footprints in system logs (e.g., power outage).
Several false negatives were also caused by calculating the wrong threshold for the SG parameter (e.g. shortly after major system updates).
Fine tuning the threshold calculator -via comparing similar vicinities- improves the recall rate by about $4\%$.

\section{Related Work}
\label{sec:relatedwork}

Failures in traditional high performance computing systems were either completely ignored, avoided or at most addressed in the middleware layer\cite{Rajachandrasekar2012}.
The advances in the size and complexity of the HPC systems, demanded more flexible approaches against failures since they are a norm rather than an exception\cite{Patterson2002}.
The checkpoint/restart and redundancy became the de-facto methods to protect HPC systems against node failures.
However, both methods impose performance and energy consumption penalties.
Knowing the point of failure and adapting the checkpointing period or the redundancy level accordingly will increase performance and reduce energy consumption significantly\cite{Dauwe2017}.
Despite the existence of failure protection mechanisms, the ever decreasing mean time between failures remains a key challenge for Exascale computing systems\cite{Geist2017}.

Considering the high number of components in modern HPC systems and the strong correlations between node failures \cite{Sayed2013,Ghaisvand2016a}, several studies investigate behavioral analysis to predict failures via anomaly detection.
Both supervised\cite{Nakka2011} and unsupervised\cite{Oliner2008} approaches were proposed.
Additionally, several tools were designed to assist in the detection of node failure correlations\cite{Di2017,vaarandi2015}.
Many of the proposed approaches, such as proactive failure detection\cite{kimura2015} are limited to certain components of the system.
More general approaches, such as a framework for detecting anomalous messages \cite{vaarandi2018} requires access to the full text of the original Syslog entries.

All the above-mentioned works are based on system log analysis.
The huge volume of system logs generated by HPC systems demands more automated approaches that take advantage of big data analysis\cite{Park2017} and machine learning algorithms.
Algorithms, models and tools such as those proposed by Anwesha\cite{Anwesha2018}, Zhang\cite{Zhang2016}, Gupta\cite{Gupta2017}, and Du\cite{Min2017} are few examples of using deep learning for predicting failures in HPC systems via Syslog analysis.

The drawback of all application-dependent approaches is the overhead incurred to the HPC system.
Using high levels of redundancy and special purpose designed architectures are also not general solutions for existing and operational HPC systems.
In addition, the available failure prediction methods are either using models configured for a specific HPC system or require detailed information about all events occurring on the HPC system which endangers users' privacy.
A working solution should be applicable to existing operational HPC systems and should not incur extensive overhead to the underlying system.
Furthermore, to the best of our knowledge, none of the existing approaches consider users' privacy protection as a fundamental principle in their analysis.

The proposed approach in this work employs a statistical Syslog analysis to detect anomalies in existing and operational HPC systems.
In contrast to other approaches, the proposed approach protects users' privacy by design.
Since the input data (system logs) are passively collected, always anonymized, and no modification of the original HPC system is required, the proposed approach is applicable to virtually all existing and operational HPC systems.
Furthermore, system logs can be replaced by any other persistent source of information which reflects the HPC system status.

\section{Conclusion and Future Work}
\label{sec:conclusion}

Given that failures are becoming an integral part of HPC systems, employing mechanisms that detect failures as early as possible significantly reduces the recovery costs and reduces the interruptions in the normal operation of HPC systems.
In this work, a node failure detection mechanism via anomaly detection on system logs was proposed, which calculates the nodes system log generation frequency (the SG parameter) during a fixed time interval and compares this parameter against other nodes in the vicinity.
The concept of vicinity defines new dimensions of node correlation, beyond the natural temporal and spatial correlations.
Abnormal behavior and thus a node failure is detected by reaching a deviation threshold from the SG parameter of the majority of the nodes.
Eight months of system logs collected from the Taurus HPC cluster were used as the main data source.
The proposed failure detection mechanism was applied to Taurus system logs in four different vicinities.
According to the results, the most effective vicinities were identified as \emph{physical location} and \emph{hardware architecture}.
Finally, the proposed mechanism was applied to each rack of computing nodes with similar hardware architecture within the Taurus HPC cluster.
Using each of the original or fully anonymized system log entries a node failure detection precision of $62\%$ could be reached.
Filtering out frequent common Syslog entries improved the precision of node failure detection to $87\%$.
It has been also shown that the proposed mechanism could detect $75\%$ of node failures with more than $80\%$ precision via analyzing filtered anonymized Syslog entries.

For future work, it is planned to dynamically adjust the interval of the sliding time window which calculates the SG parameter of each node, according to the feedback received from the \emph{time of failure} vicinity.
Also, at the moment, the detection occurs rather close to the time of failure, which could be improved by better filtering of the input data.
Furthermore, the resource allocation vicinity will be further studied via analyzing jobs' information that are executed across multiple nodes.
Performing similar analyses on publicly available system logs, such as those from the Failure Trace Archive (FTA)\footnote{fta.scem.uws.edu.au/index.php?n=Main.DataSets} or the Computer Failure Data Repository (CFDR)\footnote{www.usenix.org/cfdr}, as well as other types of monitoring data, such as node power consumption, are also planned as future work.

\section*{Acknowledgment}
This work is in part supported by the German Research Foundation (DFG) within the Cluster of Excellence `Center for Advancing Electronics Dresden (cfaed)', and by Eucor~--~ \mbox{The European} Campus, within the Seed Money project `Data Analysis for Improving High Performance Computing Operations and Research.' 

\begin{spacing}{1}
{\small
\Urlmuskip=0mu plus 1mu\relax
\bibliographystyle{./style/IEEEtranBST/IEEEtran}
\bibliography{literature}
}
\end{spacing}

\end{document}